\documentclass[preprintnumbers,amsmath,amssymbm,prd]{revtex4}
\usepackage{epsfig}
\usepackage{graphicx}

\begin{document}
\title{The holographic entropy bound in higher-dimensional spacetimes: As strong as ever}
\author{Shahar Hod}
\affiliation{The Ruppin Academic Center, Emeq Hefer 40250, Israel}
\affiliation{ } \affiliation{The Hadassah Institute, Jerusalem
91010, Israel}
\date{\today}

\begin{abstract}
\ \ \ The celebrated holographic entropy bound asserts that, within
the framework of a self-consistent quantum theory of gravity, the
maximal entropy (information) content of a physical system is given
by one quarter of its circumscribing area: $S\leq
S_{\text{max}}={\cal A}/4{\ell^2_P}$ (here $\ell_P$ is the Planck
length). An intriguing possible counter-example to this fundamental
entropy bound, which involves {\it homogenous} weakly
self-gravitating confined thermal fields in higher-dimensional
spacetimes, has been proposed almost a decade ago. Interestingly, in
the present paper we shall prove that this composed physical system,
which at first sight seems to violate the holographic entropy bound,
actually conforms to the entropy-area inequality $S\leq {\cal
A}/4{\ell^2_P}$. In particular, we shall explicitly show that the
homogeneity property of the confined thermal fields sets an upper
bound on the entropy content of the system. The present analysis
therefore resolves the apparent violation of the holographic entropy
bound by confined thermal fields in higher-dimensional spacetimes.
\end{abstract}
\bigskip
\maketitle


\section{Introduction}

The holographic principle \cite{Hoof,Suss,Bek2} provides a
remarkably simple geometric answer to the following physically
interesting question: What is the maximal entropy (information)
content of a spatially bounded physical system? In particular, the
holographic entropy bound asserts that the remarkably compact
inequality \cite{Hoof,Suss,Bek2,Noteunit,Notepl}
\begin{equation}\label{Eq1}
S\leq S_{\text{max}}={{\cal A}\over{4\ell^2_P}}\
\end{equation}
sets a fundamental upper bound on the entropy content of a physical
system of circumference area ${\cal A}$
\cite{Notealt,Bek4,Hodspin,BekMayHod}. Interestingly, the inequality
(\ref{Eq1}) is saturated by stationary black holes in generic
$(D+1)$-dimensional spacetimes. This fact suggests that, for a given
surface area ${\cal A}$, these fundamental solutions of the Einstein
field equations are characterized by the maximally allowed entropy:
$S_{\text{BH}}=S_{\text{max}}({\cal A})$.

It is widely believed \cite{Hoof,Suss,Bek2,Bek3} that the
holographic entropy bound (\ref{Eq1}) reflects a fundamental
physical feature of the elusive quantum theory of gravity.
Interestingly, the bound is known to be valid for
$(3+1)$-dimensional isolated physical systems with negligible
self-gravity \cite{Bek2}. One naturally wonders whether the
holographic entropy bound (\ref{Eq1}) is also respected by weakly
self-gravitating physical systems in {\it higher}-dimensional spacetimes
\cite{Bek6,Hodpb}?

This is a highly non-trivial question since one expects the entropy
content of a physical system to be an increasing function of the
number $D$ of spatial dimensions. In particular, for a
$(D+1)$-dimensional physical system with a fixed amount of energy
$E$, the larger is $D$, the more distinct ways there are to populate
the discrete energy levels of the physical system \cite{Bek6,Hodpb}.
It is therefore of physical interest to test the validity of the
holographic entropy bound (\ref{Eq1}) in higher-dimensional
spacetimes.

The main goal of the present paper is to calculate the entropy
content of bounded homogenous thermal fields in generic
$(D+1)$-dimensional spacetimes and to test the validity of the
fundamental holographic entropy bound (\ref{Eq1}) in these
higher-dimensional spacetimes.

\section{Entropy content of $(D+1)$-dimensional bounded thermal fields}

In a highly interesting paper, Bekenstein \cite{Bek6} has calculated
the entropy content of a $(D+1)$-dimensional physical system which
is composed of massless homogenous thermal fields of total energy
$E$ which are bounded inside a spherical box of proper radius $R$.
As explicitly shown in \cite{Bek6}, the thermal entropy content of
this physical system is given by the functional relation
\begin{equation}\label{Eq2}
S=C_D(1+1/D)N^{1\over{D+1}}{(RE/\hbar)}^{D\over{D+1}}\  ,
\end{equation}
where
\begin{equation}\label{Eq3}
C_D\equiv\Big[{{2\zeta(D+1)\Gamma({{D+1}\over{2}})}\over{\pi^{1/2}\Gamma({D\over
2})}}\Big]^{{1}\over{D+1}}\ ,
\end{equation}
and the dimensionless parameter $N$ is the effective number of
polarization states (degrees of freedom) of the higher-dimensional
thermal fields: a massless scalar field contributes $1$ to $N$, a
massless fermionic field in a $(D+1)$-dimensional spacetime
contributes $1-2^{-D}$ to $N$ \cite{Bek6}, an electromagnetic field
in a $(D+1)$-dimensional spacetime contributes $D-1$ to $N$
\cite{CarCavHal}, and the graviton in a $(D+1)$-dimensional
spacetime contributes $(D+1)(D-2)/2$ to $N$ \cite{CarCavHal}. It is
important to emphasize the fact that the entropy formula (\ref{Eq2})
was derived in Ref. \cite{Bek6} for the specific case of {\it
homogenous} massless fields in {\it thermal} equilibrium [The
homogeneity assumption made in \cite{Bek6} in deriving the entropy
formula (\ref{Eq2}) corresponds to the assumption (\ref{Eq19}) made
below].

As discussed in \cite{Bek6,Hodpb}, the compact analytical expression
(\ref{Eq2}) derived in \cite{Bek6} for the entropy content of the
$(D+1)$-dimensional thermal fields is valid under the following two
physical assumptions:
\newline
(1) The bounded physical system is {\it weakly} self-gravitating. As
explicitly shown in \cite{Hodpb}, this physical condition can be
characterized by the dimensionless strong inequality
\begin{equation}\label{Eq4}
\eta\equiv {{16\pi RE}\over{(D-1){\cal A}}}\ll 1\  ,
\end{equation}
where ${\cal A}$ is the $(D-1)$-dimensional surface area of the
bounded system.
\newline
(2) In addition, the analytical expression (\ref{Eq2}) for the
entropy content of the bounded fields is only valid in the
thermodynamic regime. As discussed in \cite{Bek6,Hodpb}, the
thermodynamic (continuum) treatment of the confined fields is valid
provided many field quanta are thermally excited. In particular, as
explicitly shown in \cite{Hodpb}, this physical condition can be
characterized by the dimensionless strong inequality
\begin{equation}\label{Eq5}
\xi\equiv
{{D}\over{C_D}}{\Big({{N\hbar}\over{RE}}\Big)}^{D\over{D+1}}\ll 1\ .
\end{equation}

Substituting (\ref{Eq4}) and (\ref{Eq5}) into (\ref{Eq2}), one finds
\begin{equation}\label{Eq6}
S=\Big({{C_D}\over{D}}\Big)^{{D+1}\over{D}}{{D^2-1}\over{4\pi}}\eta\xi^{{1}\over{D}}\cdot
{{{\cal A}}\over{4\hbar}}\
\end{equation}
for the thermal entropy of the bounded $(D+1)$-dimensional fields.
Intriguingly, the entropy-area relation (\ref{Eq6}) suggests, at
first sight, that in the large-$D$ regime \cite{Hodpb,Notecd,Noteeb}
\begin{equation}\label{Eq7}
D>{{4\pi}\over{\eta}}\gg1\  ,
\end{equation}
the entropy content of this higher-dimensional physical system can
violate the holographic entropy bound (\ref{Eq1}) \cite{Hodpb}.

As discussed above, the holographic entropy bound (\ref{Eq1}) is
widely believed \cite{Hoof,Suss,Bek2,Bek3} to reflect a fundamental
physical property of the elusive quantum theory of gravity. Thus,
for the advocates of the holographic principle, the task remains to
find out how the holographic entropy bound (\ref{Eq1}) eventually
survives the serious challenge which is imposed [see Eqs.
(\ref{Eq6}) and (\ref{Eq7})] by the higher-dimensional confined
thermal fields.

The main goal of the present paper is to explore the (in)validity of
the holographic entropy bound (\ref{Eq1}) in higher-dimensional
physical systems which are made of confined thermal fields. As we
shall show in the next section, the physical assumption of
homogeneity made in \cite{Bek6,Hodpb} sets an upper bound on the entropy content of the
fields. In particular, below we shall explicitly prove that,
properly taking into account the homogeneity property of the
confined fields, one finds that the higher-dimensional thermal
system actually respects the holographic entropy bound (\ref{Eq1}).

\section{The homogeneity assumption and the holographic
entropy bound}

In the present section we shall show that the homogeneity assumption
of the confined fields, on which the entropy formula (\ref{Eq2}) of
\cite{Bek6} is based, sets an upper bound on the dimensionless
physical quantity $RE/\hbar$ which characterizes the spatially
bounded $(D+1)$-dimensional physical system. In particular, we shall explicitly prove
that, for higher-dimensional spacetimes, the new bound on the
dimensionless quantity $RE/\hbar$, which stems from the homogeneity
assumption [see Eq. (\ref{Eq23}) below], is stronger than the
formerly used bound (\ref{Eq4}) which stems from the weak gravity
assumption.

The line element of a curved $(D+1)$-dimensional spherically
symmetric spacetime can be expressed in the form
\begin{equation}\label{Eq8}
ds^2=e^{\nu}dt^2-e^{\lambda}dr^2-r^2d\Omega\  ,
\end{equation}
where $\nu=\nu(r)$ and $\lambda=\lambda(r)$ are dimensionless metric
functions and $d\Omega$ is the angular line element of a unit
$(D-1)$-dimensional sphere. Assuming an isotropic matter
distribution whose energy momentum tensor is given by
$T^{\mu}_{\nu}={\text{diag}}(\rho,-p,-p,...,-p)$ \cite{Noterp}, the
$(D+1)$-dimensional Einstein field equations,
$R_{\mu\nu}=8\pi[T_{\mu\nu}-{{1}\over{D-1}}g_{\mu\nu}T]$, take the
form \cite{Leon}
\begin{equation}\label{Eq9}
e^{-\lambda}\Big({{\lambda'}\over{r}}-{{D-2}\over{r^2}}\Big)+{{D-2}\over{r^2}}={{16\pi}\over{D-1}}\rho\
,
\end{equation}
and
\begin{equation}\label{Eq10}
e^{-\lambda}\Big({{\nu'}\over{r}}+{{D-2}\over{r^2}}\Big)-{{D-2}\over{r^2}}={{16\pi}\over{D-1}}p\
.
\end{equation}

The mass (energy) contained within a $D$-dimensional sphere of
radius $r$ is given by \cite{SchTang,Kun}
\begin{equation}\label{Eq11}
E(r)=A_{D-1}\int^{r}_{0}\rho(r')r'^{D-1}dr'\ ,
\end{equation}
where
\begin{equation}\label{Eq12}
A_{D-1}={{2\pi^{D/2}}\over {\Gamma(D/2)}}
\end{equation}
is the surface area of a unit $(D-1)$-dimensional sphere. From Eq.
(\ref{Eq9}) one finds the simple relation \cite{Leon}
\begin{equation}\label{Eq13}
e^{-\lambda}=1-{{2{\bar E}(r)}\over{r^{D-2}}}\  ,
\end{equation}
where
\begin{equation}\label{Eq14}
{\bar E}(r)\equiv{{8\pi}\over{(D-1)A_{D-1}}}E(r)\ .
\end{equation}
Substituting Eq. (\ref{Eq13}) into Eq. (\ref{Eq10}), one obtains the
differential relation \cite{Leon}
\begin{equation}\label{Eq15}
\nu'={{2}\over{D-1}}\cdot{{(D-1)(D-2){\bar E}(r)+8\pi
pr^D}\over{r^{D-1}[1-2{\bar E}(r)/r^{D-2}]}}\  .
\end{equation}

In addition, substituting the Einstein field equations (\ref{Eq9})
and (\ref{Eq10}) into the conservation equation
\begin{equation}\label{Eq16}
T^{\mu}_{r;\mu}=0\  ,
\end{equation}
one finds the differential relation \cite{Leon}
\begin{equation}\label{Eq17}
{{dp}\over{dr}}=-{1\over2}\nu'(\rho+p)\
\end{equation}
for the gradient of the radial pressure. Taking cognizance of Eqs.
(\ref{Eq15}) and (\ref{Eq17}), one obtains the generalized
[$(D+1)$-dimensional] Tolman-Oppenheimer-Volkov equation
\begin{equation}\label{Eq18}
{\cal F}\equiv
{{r}\over{p}}\cdot{{dp}\over{dr}}=-{{(1+\rho/p)[(D-1)(D-2){\bar
E}(r)+8\pi p r^D]}\over{(D-1)r^{D-2}[1-2{\bar E}(r)/r^{D-2}]}}\  .
\end{equation}

At this point, it is important to emphasize again that the expression
(\ref{Eq2}) for the entropy content of the higher-dimensional
physical system is valid for weakly self-gravitating confined
thermal fields of spatially homogeneous ({\it constant}) energy density and pressure
\cite{Bek6}. This homogeneity assumption made in \cite{Bek6,Hodpb} corresponds to the
dimensionless strong inequality [see Eq. (\ref{Eq18})]
\begin{equation}\label{Eq19}
{\cal F}\ll1\  .
\end{equation}
Taking cognizance of Eqs. (\ref{Eq18}) and (\ref{Eq19}), and using
the relation \cite{Hav}
\begin{equation}\label{Eq20}
p=\rho/D\
\end{equation}
for massless radiation fields in $D$ spatial dimensions,
one finds the characteristic strong inequality \cite{Noteapr}
\begin{equation}\label{Eq21}
(D-2)(D+1){{{\bar E}(R)}\over{R^{D-2}}}\ll1\
\end{equation}
for the nearly homogenous entropy-bearing thermal fields.
Substituting (\ref{Eq14}) into (\ref{Eq21}), one obtains the strong inequality
\begin{equation}\label{Eq22}
{{8\pi(D-2)(D+1)}\over{(D-1)A_{D-1}}}\cdot{{E}\over{R^{D-2}}}\ll1\ ,
\end{equation}
which yields the characteristic small dimensionless ratio
\cite{Noteaer}
\begin{equation}\label{Eq23}
\epsilon\equiv {{8\pi(D-2)(D+1)}\over{D-1}}\cdot{{RE}\over{{\cal
A}}}\ll1\
\end{equation}
for the nearly homogenous confined thermal fields.

Finally, substituting (\ref{Eq5}) and (\ref{Eq23}) into (\ref{Eq2}),
one obtains the functional expression
\begin{equation}\label{Eq24}
S=\alpha(D)\cdot{{\xi^{{{1}\over{D}}}\epsilon{\cal
A}}\over{4\hbar}}\
\end{equation}
for the entropy of the confined $(D+1)$-dimensional {\it homogenous}
thermal fields, where
\begin{equation}\label{Eq25}
\alpha(D)\equiv{{{(D-1)C^{{{D+1}\over{D}}}_D}\over{2\pi(D-2)D^{{D+1}\over{D}}}}}\
.
\end{equation}
In Table \ref{Table1} we present the numerically computed values of
the $D$-dependent pre-factor $\alpha(D)$. The numerically computed
values of the pre-factor reveal the fact that $\alpha (D)$ is a
monotonically decreasing function of the number $D$ of spatial
dimensions with the property
$\text{max}_{D}\{\alpha(D)\}=\alpha(D=3)\simeq0.0819$.

\begin{table}[htbp]
\centering
\begin{tabular}{|c|c|c|c|c|c|c|c|c|c|}
\hline $D$ & \ $\ 3\ $\ \ & \ $\ 4\ $\ \ & \
$\ 5\ $\ \ \ & \ $\ 6\ $\ \ & \ $\ 7\ $\ \ & \ $\ 8\ $\ \ & \ $\ 9\ $\ \ & \ $\ 10\ $\ \\
\hline \ $\ \ \alpha(D)\ $\ \ \ &\ \ $0.0819$\ \ \ &\ \ $0.0471$\ \
\ &\ \ $0.0343$ \ \ \ &\ \ $0.0274$\ \ \ &\ \ $0.0229$\ \ \ &\ \
$0.0197$\ \ \ &\
\ $0.0174$\ \ \ &\ \ $0.0156$\ \ \ \\
\hline
\end{tabular}
\caption{The pre-factor $\alpha(D)$ of the entropy-area relation
(\ref{Eq24}). One finds that $\alpha (D)$ is a monotonically
decreasing function of the number $D$ of spatial dimensions with the
property $\text{max}_{D}\{\alpha(D)\}=\alpha(D=3)\simeq0.0819$.
Taking cognizance of Eq. (\ref{Eq24}), one deduces that the
$(D+1)$-dimensional homogenous thermal fields conform to the
holographic entropy bound (\ref{Eq1}).} \label{Table1}
\end{table}

Interestingly, taking cognizance of the strong dimensionless
inequalities (\ref{Eq5}) and (\ref{Eq23}) and the data presented in
Table \ref{Table1}, one deduces that, for generic values of the
number $D$ of spatial dimensions, the entropy content (\ref{Eq24})
of the $(D+1)$-dimensional thermal fields actually {\it conforms} to
the holographic entropy bound (\ref{Eq1}).

\section{Summary}

The holographic entropy bound \cite{Hoof,Suss,Bek2} asserts that the
maximal entropy content of a bounded physical system is given (in
Planck units) by one quarter of its circumscribing area: $S\leq
{\cal A}/4{\ell^2_P}$ [see Eq. (\ref{Eq1})]. It is widely believed
\cite{Hoof,Suss,Bek2,Bek3} that this characteristic entropy bound
reflects a fundamental physical property of the elusive quantum
theory of gravity. It is therefore quite surprising that the entropy
of homogenous weakly self-gravitating confined thermal fields in
higher-dimensional spacetimes seems, at first sight, to violate the
entropy-area relation (\ref{Eq1}) [see Eqs. (\ref{Eq6}) and
(\ref{Eq7})] \cite{Hodpb}.

Interestingly, in the present paper we have explicitly proved that
the {\it homogeneity} assumption made in \cite{Bek6,Hodpb} for
the confined thermal fields [see Eq. (\ref{Eq23})] sets an upper bound on the entropy content of the
physical system. In particular, it has been shown that,
properly taking into account the assumed homogeneity property of the spatially bounded physical system,
one finds that the higher-dimensional thermal fields
actually conforms to the holographic entropy bound
(\ref{Eq1}). Most importantly, from Eqs. (\ref{Eq23}) and (\ref{Eq24})
one obtains the asymptotic expression \cite{Notecd}
\begin{equation}\label{Eq26}
S={{\epsilon}\over{2\pi D}}\cdot{{{\cal A}}\over{4\hbar}}\ll {{{\cal
A}}\over{4\hbar}}\ \ \ \ \text{for}\ \ \ \ D\gg1
\end{equation}
for the entropy content of the higher-dimensional fields in the
large-$D$ regime. The present analysis therefore resolves the
apparent violation \cite{Hodpb} of the holographic entropy bound by
confined thermal fields in higher-dimensional spacetimes.

\bigskip
\noindent {\bf ACKNOWLEDGMENTS}
\bigskip

This research is supported by the Carmel Science Foundation. I thank
Yael Oren, Arbel M. Ongo, Ayelet B. Lata, and Alona B. Tea for
stimulating discussions.


\begin{thebibliography}{99}
\bibitem{Noteunit} We shall use natural units in which $G=c=k_B=1$.
\bibitem{Notepl} Here $\ell_{\text{P}}=\sqrt{G\hbar/c^3}$ is the characteristic Planck length.
\bibitem{Notecd} Here we have used the asymptotic relations $C_D\to 1$, $D^{1+{{1}/{D}}}\to D[1+(\ln D)/D]$,
and $\xi^{{1}\over{D}}\to1$ for $D\gg 1$.
\bibitem{Noteeb} The strong inequality on the r.h.s of (\ref{Eq7}) stems from the
fact that $\eta\ll1$ [see Eq. (\ref{Eq4})].
\bibitem{Noterp} Here $\rho$ is the energy density and $p$ is the
corresponding pressure component.
\bibitem{Noteapr} Here we have used the relation $p>0$ which
implies $(D-1)(D-2){\bar E}(r)+8\pi p r^D>(D-1)(D-2){\bar E}(r)$. In
addition, we have used here the inequality $1-2{\bar
E}(r)/r^{D-2}<1$ for the spatially regular configurations.
One therefore finds $|{\cal F}|>{{(1+\rho/p)(D-2){\bar E}(r)}\over{r^{D-2}}}$.
\bibitem{Noteaer} Here we have used the relation ${\cal A}=A_{D-1}R^{D-1}$ for the $(D-1)$-dimensional
surface area of the bounded spherical system.
\end{thebibliography}
\end{document}